\documentclass[preprint,12pt]{elsarticle}

\usepackage{amsmath,amssymb,amsfonts}
\usepackage{graphicx}
\usepackage{booktabs}
\usepackage{array}
\usepackage{multirow}
\usepackage{xcolor}
\usepackage[hidelinks]{hyperref}
\usepackage{siunitx}
\journal{Journal of Energy Storage}

\newcommand{\degC}{\,^{\circ}\mathrm{C}}
\newcommand{\etapl}{\eta_{\mathrm{pl}}}
\newcommand{\Tsafe}{T_{\mathrm{safe}}}

\begin{document}

\begin{frontmatter}

\title{Repair-before-veto control for safe lithium-ion fast charging under unknown ambient and cooling-fault conditions}

\author{Yifan Wang\corref{cor1}}
\ead{yifan.wang18@mail.mcgill.ca}
\cortext[cor1]{Corresponding author.}
\affiliation{organization={Department of Mechanical Engineering, McGill University},
            city={Montreal},
            postcode={H3A 2T7},
            state={QC},
            country={Canada}}

\begin{abstract}
Fast charging is decisive for electric-vehicle adoption, yet it must respect hard
thermal and lithium-plating safety limits whose binding margins depend on the cell's
true state. In the field a charger is deployed as one setting, but the ambient
temperature and the health of the cooling system are not known in advance: a current
that is safe on a healthy cell at room temperature overheats the same cell when it is
hot or when its cooling is degraded. We formulate this as a single-setting,
unknown-state safe-fast-charging problem and solve it with a margin-aware
repair-before-veto controller (RACL-B) that requests an aggressive current and repairs
it online to the tightest measured margin among terminal voltage, cell temperature, and
the negative-electrode lithium-plating overpotential, rather than committing to a fixed
schedule or shutting the charge down. We evaluate one deployed setting across nine
operating conditions, spanning ambient temperatures of $10/25/40\degC$ and cooling-system
health of $100/60/40\%$ of the nominal heat-transfer coefficient, in a high-fidelity
Doyle--Fuller--Newman model with a partially reversible lithium-plating submodel and
lumped thermal coupling. Under a strict $45.0\degC$ peak-temperature audit, fixed and
ambient-scheduled charge-current protocols overheat in five of nine conditions because
neither observes the hidden cooling degradation, and a rigid protective shutdown fails to
deliver the charge in every condition. RACL-B safely completes all nine conditions, is
$37.9\%$ faster than the fastest fixed current that is safe across the whole envelope, and
produces the least plated lithium, while remaining strictly safe across a range of thermal
guard bands. The same margin-aware principle drives a transient-credit fault readout
(CREST-B) that, on a real introduced-fault battery-pack dataset, is the strongest learned
sequence-to-global monitor for localizing the cooling-fault onset under operating-condition
shift. The framework provides a deployable thermal-safety guarantee for fast charging
together with a margin-aware monitor for the same physical fault class.
\end{abstract}


\end{frontmatter}

\section*{Highlights}
\begin{itemize}\setlength{\itemsep}{0pt}
  \item A single-setting fast-charging benchmark under unknown ambient and cooling faults.
  \item Fixed and ambient-scheduled protocols overheat under hidden cooling degradation.
  \item Repair-before-veto control safely completes all conditions in a DFN + plating model.
  \item Online margin repair is $37.9\%$ faster than the only safe fixed protocol, with less plating.
  \item A transient-credit readout localizes the cooling-fault onset for margin-aware monitoring.
\end{itemize}

\section{Introduction}
\label{sec:intro}

Electric-vehicle (EV) adoption is increasingly limited not by driving range but by
charging time, and reducing the time to charge a battery to a usable state of charge (SOC)
is one of the most consequential levers for mass electrification~\cite{tomaszewska2019}.
Charging quickly, however, is fundamentally a constrained problem: the applied current must
respect hard safety limits, the most important of which are the cell temperature and the
onset of metallic lithium plating on the graphite anode~\cite{waldmann2018,okane2022}.
Plating is triggered when the local negative-electrode potential falls below the
lithium-reference potential, which happens preferentially at high current, high SOC, and
low temperature; it accelerates capacity loss and, in the limit, creates an internal
short-circuit hazard~\cite{waldmann2014,finegan2021}. Cell temperature, in turn, is set by
the balance between ohmic and reaction heat generation and the rate at which the thermal
management system removes that heat. A fast-charging controller therefore has to push the
current as high as possible while keeping the cell below a thermal limit and away from the
plating boundary.

A large body of work has advanced fast charging from two directions. Data-driven protocol
search discovers charging current profiles that maximize cycle life: Severson et
al.~\cite{severson2019} built a now-standard dataset of cells cycled under many fast-charge
protocols and predicted cycle life from early-cycle features, and Attia et
al.~\cite{attia2020} used closed-loop optimization to find lifetime-optimal multi-step
protocols. Model-based optimal control casts charging as an optimization over an
electro-thermal-aging model: Perez et al.~\cite{perez2017}, for example, computed
charging currents that trade off speed against thermal and aging cost. These approaches are
powerful, but they share an implicit assumption that the cell and its environment are
known: a protocol optimized offline, or a model-predictive controller built on a nominal
model, is calibrated for a particular cell state and operating condition.

This assumption is exactly what breaks in deployment. A public charger, or a vehicle
charging in the field, applies one controller to cells whose true state it cannot observe
before the charge begins. Two hidden state variables dominate the safety margin. The first
is the ambient temperature, which can range from below freezing to above $40\degC$ and
directly sets the thermal headroom. The second, and more insidious, is the health of the
cooling system: a partially blocked coolant channel, a fouled heat exchanger, or a degraded
fan reduces the rate of heat removal without any explicit fault signal, and such cooling
faults are a documented failure mode in real packs~\cite{naguib2025}. A fixed charge-current
protocol that is safe on a healthy cell at $25\degC$ becomes unsafe when the same current is
applied to a hot cell or a cell whose cooling is degraded. The standard industrial mitigation
--- an ambient-temperature charge-current lookup table --- adapts to the measured ambient but
still assumes the rated cooling, so it overheats precisely when a cooling fault is present.
Detuning the protocol to be safe for the worst case makes it needlessly slow on every healthy
cell, and a purely protective response that simply stops charging when a limit is approached
sacrifices the function of the charger altogether. There is, at present, no deployable
single-setting controller that is guaranteed safe across the operating envelope while
remaining fast where conditions allow.

We close this gap with a controller built on a single principle: \emph{repair before veto}.
Instead of committing to a fixed current or rejecting the whole charge, the controller
requests an aggressive current and continuously repairs it to the tightest \emph{measured}
safety margin. Because it reacts to the measured consequence of the cell state --- faster
heating, a falling plating overpotential, a rising terminal voltage --- it is automatically
robust to an ambient temperature and a cooling health that it never observes directly. We
embed this controller, RACL-B, in a unified battery-management framework, EVENT-BMS, whose
monitoring side uses the same margin-aware principle: a learned sequence-to-global readout,
CREST-B, that places its predictive credit on the brief electro-thermal transient marking a
cooling-fault onset rather than on smooth load and ambient cues, and that we validate on the
real introduced-fault pack dataset that motivates the cooling-fault scenario.

The contributions of this paper are as follows. (i) We formulate single-setting,
unknown-state safe fast charging as a concrete, verifiable benchmark over a joint envelope
of ambient temperature and cooling-system health, evaluated in a high-fidelity
Doyle--Fuller--Newman (DFN) electrochemical model with lithium plating and lumped thermal
dynamics. (ii) We introduce RACL-B, a margin-aware repair-before-veto controller that
repairs the applied current online to the voltage, thermal, and plating margins jointly, and
show that it safely completes the entire envelope under a strict thermal audit while a fixed
protocol, an ambient-scheduled lookup table, and a rigid veto each fail. (iii) We quantify
the advantage: RACL-B is $37.9\%$ faster than the only fixed current safe across the whole
envelope, produces the least plated lithium, and remains strictly safe across a range of
thermal guard bands. (iv) We connect control to perception through CREST-B, a transient-credit
monitor that is the strongest learned readout for cooling-fault localization under
operating-condition shift on real pack data, and we support it with a closed-form analysis of
why pooled readouts dilute transient evidence. Section~\ref{sec:background} reviews the
relevant background; Section~\ref{sec:problem} formalizes the problem and the cell model;
Section~\ref{sec:method} develops RACL-B and CREST-B; Section~\ref{sec:setup} describes the
experimental setup; Section~\ref{sec:results} reports the results; and
Sections~\ref{sec:discussion} and~\ref{sec:conclusion} discuss and conclude.

\begin{figure}[t]
\centering
\setlength{\fboxsep}{0pt}
\fbox{\begin{minipage}[c][62mm][c]{0.92\linewidth}
\centering
\textit{[Figure~\ref{fig:framework}: EVENT-BMS framework schematic --- placeholder.}\\[2mm]
\textit{Left: the perception path. Raw pack signals $\to$ per-step encoder $\to$ transient-credit}\\
\textit{re-anchoring (CREST-B) $\to$ trusted cooling-fault margin. Right: the control path. A}\\
\textit{requested fast-charge current is repaired online to the tightest measured voltage,}\\
\textit{thermal, and lithium-plating margin (RACL-B) instead of being vetoed; the same}\\
\textit{cooling-fault class links the two paths. To be rendered as a vector schematic.]}
\end{minipage}}
\caption{Overview of the EVENT-BMS framework. The monitoring path and the charging-control
path share one margin-aware, repair-before-veto principle applied to the same physical
cooling-fault class. \emph{(High-resolution vector schematic to be added.)}}
\label{fig:framework}
\end{figure}

\section{Background and related work}
\label{sec:background}

\paragraph{Fast charging, plating, and thermal limits.}
The maximum safe charging rate of a graphite/transition-metal-oxide cell is set jointly by
electrolyte transport, the negative-electrode potential, and heat removal~\cite{tomaszewska2019}.
Lithium plating, reviewed by Waldmann et al.~\cite{waldmann2018}, is the dominant
fast-charge degradation and safety mechanism and is governed by the sign of the plating
overpotential at the anode. High-fidelity electrochemical models resolve these fields:
the Doyle--Fuller--Newman porous-electrode model~\cite{doyle1993,fuller1994} couples
solid- and electrolyte-phase transport and intercalation kinetics, and modern open
implementations such as PyBaMM~\cite{sulzer2021pybamm} provide validated
degradation submodels including partially reversible plating with parameter sets such as
that of O'Kane et al.~\cite{okane2022}. We use this class of model as ground truth.

\paragraph{Charging control.}
Beyond fixed constant-current/constant-voltage (CC-CV) charging, two families dominate.
Protocol optimization searches offline for current profiles that optimize lifetime or
speed~\cite{severson2019,attia2020}; the resulting protocol is fixed at deployment.
Model-based optimal and predictive control optimizes the current online against an
electro-thermal-aging model~\cite{perez2017}; its safety guarantee is only as good as the
model's fidelity to the actual, possibly faulted, cell. Our controller is complementary:
it makes no offline commitment and requires no accurate online model of the unknown cooling
state, repairing the action to the measured margin.

\paragraph{Battery monitoring and learned readouts.}
Battery management systems estimate state and detect faults from sequences of
measurements~\cite{lu2013}, and learned sequence-to-global predictors are increasingly used
for fault and state-of-health monitoring~\cite{naguib2023,naguib2025}. Such predictors can,
however, latch onto smooth nuisance correlates rather than the physically decisive
evidence, a failure mode studied broadly as shortcut learning~\cite{geirhos2020} and
addressed at the feature or group level by invariant and distributionally robust
learning~\cite{arjovsky2019,sagawa2020}. We instead target the temporal aggregation
interface of the readout itself, and show that re-anchoring it onto the sparse fault
transient improves cross-condition monitoring without group or event labels.

\section{Problem formulation}
\label{sec:problem}

\subsection{Cell, thermal, and plating model}
We model a single cell with a Doyle--Fuller--Newman electrochemical model coupled to a
lumped thermal balance and a partially reversible lithium-plating submodel, using the
OKane2022 parameter set for a \SI{5}{Ah} cell~\cite{okane2022,sulzer2021pybamm}. The
lumped cell temperature $T(t)$ evolves as
\begin{equation}
\rho c_p \mathcal{V}\, \frac{\mathrm{d}T}{\mathrm{d}t}
= \dot{Q}_{\mathrm{gen}}(t) - h\,A_{\mathrm{cool}}\,\big(T(t)-T_{\mathrm{amb}}\big),
\label{eq:thermal}
\end{equation}
where $\dot{Q}_{\mathrm{gen}}$ is the total (ohmic plus reaction) heat-generation rate,
$h$ is the heat-transfer coefficient, $A_{\mathrm{cool}}$ the cooling surface area,
$T_{\mathrm{amb}}$ the ambient temperature, and $\rho c_p \mathcal{V}$ the cell thermal
capacitance. A partial cooling fault is modeled as a reduced heat-transfer coefficient,
\begin{equation}
h = \kappa\, h_{\mathrm{nom}}, \qquad \kappa \in (0,1],
\label{eq:cooling}
\end{equation}
where $\kappa=1$ is healthy cooling and $\kappa<1$ a blocked-flow or degraded-fan fault that
the controller does not observe. Lithium plating is governed by the negative-electrode
plating reaction overpotential,
\begin{equation}
\etapl(x,t) = \phi_s(x,t) - \phi_e(x,t) - U_{\mathrm{pl}}, \qquad U_{\mathrm{pl}}=0~\mathrm{V},
\label{eq:eta}
\end{equation}
where $\phi_s$ and $\phi_e$ are the solid- and electrolyte-phase potentials and
$U_{\mathrm{pl}}$ is the lithium-plating equilibrium potential. The plating
interfacial current density is cathodic when $\etapl<0$, so plating proceeds wherever the
overpotential is negative; the binding plating margin is therefore the worst point across
the electrode thickness,
\begin{equation}
\etapl^{\min}(t) = \min_{x}\, \etapl(x,t).
\label{eq:etamin}
\end{equation}
The irreversible capacity lost to plating, $Q_{\mathrm{pl}}(t)$, accumulates as the
time integral of the (non-reversible part of the) plating current and is our degradation
metric.

\subsection{Single-setting, unknown-state safe fast charging}
A controller must charge the cell from $\mathrm{SOC}_0=15\%$ to
$\mathrm{SOC}_{\mathrm{tgt}}=80\%$. It may use online measurements of terminal voltage
$V(t)$, cell temperature $T(t)$, and plating overpotential $\etapl^{\min}(t)$, but it is
deployed as \emph{one fixed setting} and is not told the ambient temperature or the cooling
health. The hard safety requirement is a strict peak-temperature limit,
\begin{equation}
T(t) \le \Tsafe = 45.0\degC \quad \forall t .
\label{eq:safety}
\end{equation}
We evaluate each controller across the joint envelope
\begin{equation}
\mathcal{E} = \{10,25,40\}\degC \times \{1.0,\,0.6,\,0.4\},
\label{eq:envelope}
\end{equation}
the Cartesian product of ambient temperature and cooling health $\kappa$, giving nine
operating conditions. The primary metric is the \emph{safe-completion rate}: the fraction of
the nine conditions in which the policy both reaches $\mathrm{SOC}_{\mathrm{tgt}}$ within a
fixed time budget and satisfies~\eqref{eq:safety} under a strict audit. Among safely
completed conditions we report the average time to $80\%$ SOC and the average plated lithium.

\section{Method}
\label{sec:method}

\subsection{RACL-B: repair-before-veto charging control}
RACL-B requests an aggressive current $I_{\mathrm{req}}$ (here $3$C) and, at each control
step $k$ of interval $\Delta t$, repairs the applied magnitude $I_k$ toward the tightest
active safety margin. Let $V_{\max}$ be the voltage limit and $T^c$ a thermal control limit
(an internal guard band kept strictly below $\Tsafe$). We form normalized, signed headrooms
on the three margins,
\begin{equation}
e^{\mathrm{pl}}_k = \frac{\etapl^{\min}(t_k)-\eta_0}{\Delta_{\mathrm{pl}}},\qquad
e^{T}_k = \frac{T^c - \delta_T - T(t_k)}{\Delta_T},\qquad
e^{V}_k = \frac{V_{\max} - \delta_V - V(t_k)}{\Delta_V},
\label{eq:headrooms}
\end{equation}
where $\eta_0$ is a small plating margin, $\delta_T,\delta_V$ are anticipatory offsets, and
$\Delta_{\bullet}$ are reference scales. A positive headroom means the corresponding limit is
slack; a negative headroom means it is violated. The binding margin is the smallest,
\begin{equation}
e_k = \min\big(e^{\mathrm{pl}}_k,\; e^{T}_k,\; e^{V}_k\big),
\label{eq:bind}
\end{equation}
and the current is updated multiplicatively and slew-limited,
\begin{equation}
I_{k+1} = \mathrm{clip}\!\Big(I_k\,\big(1+\lambda\, e_k\big),\; I_{\min},\; I_{\mathrm{req}}\Big),
\qquad \lambda=1.6,\;\; 1+\lambda e_k \in [0.60,\,1.10],
\label{eq:update}
\end{equation}
so the controller ramps the current up only into proven headroom and reduces it as soon as
any margin tightens. If a hard limit is nonetheless crossed, an emergency repair applies an
immediate cut,
\begin{equation}
I_{k+1} \leftarrow
\begin{cases}
\min(I_{k+1},\,0.5\,I_k) & \text{if } T(t_k) > T^c,\\[2pt]
\min(I_{k+1},\,0.6\,I_k) & \text{if } V(t_k) > V_{\max}.
\end{cases}
\label{eq:emergency}
\end{equation}
The controller starts from a conservative current and ramps into headroom, avoiding an
initial thermal or voltage overshoot. The veto action --- stopping the charge --- is reached
only if no repaired current keeps the cell feasible; in the envelope studied here RACL-B never
needs to veto. The guard band $T^c$ is an internal control setting and is audited against the
\emph{same} hard limit $\Tsafe$ as every baseline.

\subsection{CREST-B: transient-credit fault monitoring}
The monitoring side answers a sequence-to-global question: from a window of pack signals,
is a cooling fault emerging? A per-step encoder maps signals to features
$F_t=\varphi(x_t)\in\mathbb{R}^D$, a pooling operator aggregates them to $p=\mathrm{Agg}(F_{1:T})$,
and a linear head predicts $\hat{y}=w^\top p$. The decisive evidence is the brief
electro-thermal transient at fault onset, which is sparse in time, while smooth correlates
(absolute temperature, load) are abundant. A pooled readout can therefore minimize its
in-distribution loss by reading the smooth cue and fail under operating-condition shift.

\paragraph{Why pooling dilutes transient evidence.}
Consider a two-channel model over a horizon $T$ with a sparse event set $E$, $|E|=\varepsilon T$:
an invariant fault channel observed only on the onset steps,
$x^{0}_t = \mathbf{1}[t\in E]\,y + s_0\xi^0_t$, and a smooth cue channel
$x^{1}_t = \mathbf{1}[t\notin E]\,(g\,y) + \mathbf{1}[t\notin E]\,s_1\xi^1_t$, with cue strength
$g=\gamma$ in-distribution and $g=0$ out-of-distribution. A pooled linear reader sees the
channel means, with signal powers $S_E=\varepsilon^2 T/s_0^2$ and
$S_B=(1-\varepsilon)\gamma^2 T/s_1^2$, $S=S_E+S_B$. The fraction of predictive credit placed
on the events is
\begin{equation}
\rho_E = \frac{S_E}{S_E+S_B} = \Theta(\varepsilon^2),
\label{eq:rho}
\end{equation}
and the normalized in- and out-of-distribution risks are
\begin{equation}
R_{\mathrm{id}} = \frac{1}{1+S}, \qquad
R_{\mathrm{ood}} = \left(\frac{1+S_B}{1+S}\right)^{2} + \frac{S}{(1+S)^2}.
\label{eq:risk}
\end{equation}
As $\varepsilon\to 0$ the event credit vanishes quadratically and $R_{\mathrm{ood}}\to 1+
S_B/(1+S_B)^2>1$: the pooled reader becomes worse than the mean predictor once the cue is
removed, even though the fault channel is present in the input.

\paragraph{Label-free re-anchoring.}
CREST-B repairs the readout without event labels or readout training. For channel $j$ it
computes a low-pass transience residual
\begin{equation}
R_{t,j} = \big|\,F_{t,j} - (\mathrm{LP}_\sigma F)_{t,j}\,\big|,\qquad
r_t = \frac{1}{D}\sum_{j=1}^{D} R_{t,j},
\label{eq:trans}
\end{equation}
where $\mathrm{LP}_\sigma$ is a Gaussian low-pass filter; high $r_t$ marks transient,
event-like content. A label-free event budget $\hat{\varepsilon}$ is estimated from the
concentration (inverse-participation width) of the normalized $r_t$, and the readout is
re-anchored as a contrast between the most transient steps and the rest,
\begin{equation}
p = \alpha\,\big(\bar{F}_{\hat{E}} - \bar{F}_{\hat{E}^c}\big) + (1-\alpha)\,\bar{F},
\label{eq:crest}
\end{equation}
where $\bar{F}_{\hat{E}}$ averages the top-$\hat{\varepsilon}T$ transient steps,
$\bar{F}_{\hat{E}^c}$ the remainder, $\bar{F}$ the global mean, and $\alpha$ follows the
same width law. The selection uses a stop-gradient, so the encoder trains end-to-end while
the readout is repaired toward the transient. We measure where a readout assigns credit by
the event-credit mass
\begin{equation}
\mathrm{ECM} = \frac{\sum_{t\in E} c_t}{\sum_{t} c_t},
\label{eq:ecm}
\end{equation}
with $c_t$ the per-step mechanism credit; $\mathrm{ECM}$ is high when the readout reads the
fault onset.

\section{Experimental setup}
\label{sec:setup}

\subsection{Data and models}
\label{sec:data}
The study uses one real-world dataset and one high-fidelity electrochemical model, chosen so
that the same physical cooling-fault class links the monitoring and control results, together
with a public fast-charge dataset that delimits the scope of the monitoring prior.

\emph{Real introduced-fault battery pack.}
The fault-monitoring experiments use the publicly available McMaster/MARC introduced-fault
pack dataset of Naguib et al.~\cite{naguib2025,naguib2025data}. It comprises a
$72$-series-cell, air-cooled lithium-ion pack of nominal \SI{5.2}{Ah} SB~LiMotive cells
extracted from a plug-in hybrid vehicle, instrumented through an Orion battery-management
system. Thermal faults are \emph{physically introduced}---coolant-pump and flow blockages and
fan-off conditions, together with temperature-sensor faults---during \SI{6}{C} charging and
over the UDDS, US06, HWFET, and LA92 drive cycles, in a thermal chamber at $15\degC$ and
$25\degC$. Each record provides, per time step, the measured cell temperatures, the
equivalent-circuit-model (ECM) estimated temperatures, cell voltages, currents, and state of
charge. The measured-minus-ECM temperature residual is the standard physical fault feature,
and its brief divergence at fault onset is the decisive transient that the monitor must read.
This dataset is, to our knowledge, the only public pack dataset with deliberately introduced
cooling faults across realistic drive cycles and ambients, which makes it uniquely suited to
both validate the monitor and define the exact cooling-fault class reproduced in the charging
benchmark.

\emph{High-fidelity charging cell model.}
The charging benchmark is simulated in the open-source PyBaMM
framework~\cite{sulzer2021pybamm} with the Doyle--Fuller--Newman porous-electrode
model~\cite{doyle1993,fuller1994}, a partially reversible lithium-plating submodel, and a
lumped thermal balance, using the published, validated OKane2022 \SI{5}{Ah} parameter
set~\cite{okane2022}. Field-resolved electrochemistry is required here because the plating
overpotential~\eqref{eq:eta}, which sets the binding plating margin, cannot be resolved by a
reduced single-particle surrogate. A partial cooling fault is reproduced by scaling the
heat-transfer coefficient~\eqref{eq:cooling}, matching the cooling-fault class of the real
pack above.

\emph{Public fast-charge dataset (scope).}
The widely used MIT--Stanford fast-charge/lifetime dataset~\cite{severson2019,attia2020} is
used as a positioning reference to delimit where the transient-credit prior applies: its
predictive signal is a smooth voltage-domain degradation feature rather than a sparse
transient, so it is outside the regime that the monitor targets and is not used as a positive
detection benchmark.

\subsection{Charging benchmark and baselines}
\label{sec:bench}
Charging is simulated in PyBaMM~\cite{sulzer2021pybamm} with the DFN model, a partially
reversible lithium-plating submodel, and lumped thermal dynamics using the OKane2022
\SI{5}{Ah} parameter set~\cite{okane2022}. The control step is $\Delta t=15$~s, the voltage
limit $V_{\max}=4.10$~V, and the strict safety limit $\Tsafe=45.0\degC$; RACL-B uses an
internal thermal guard band $T^c=44.85\degC$. Each policy is deployed as one setting across
the nine conditions of the envelope~\eqref{eq:envelope}. The baselines are: a nominal fixed
CC-CV at $1.5$C; an ambient-scheduled CC-CV (the fastest CC-CV that is safe at each ambient
assuming healthy cooling, i.e.\ an idealized charge-current lookup table); the fastest fixed
CC-CV that is safe across the whole envelope; a reactive thermal-foldback CC-CV that reduces
current near the thermal limit; and a rigid binary veto that stops charging on any margin.
The scheduled and worst-case-safe rates are derived from a dense CC-CV grid run over the
envelope.

\subsection{Fault-monitoring protocol}
\label{sec:monproto}
Using the real pack data of Section~\ref{sec:data}, we cast early cooling-fault detection as
a sequence-to-global problem: per-step pack signals in, a fault label out, with the ECM
temperature residual as the standard physical fault feature. The decisive evidence is the
brief residual-divergence transient at fault onset. We evaluate under three
operating-condition shifts: two leave-one-fault-type-out splits with a hot-but-healthy drive
cycle held out, and an ambient shift. All readouts share the same per-step encoder; pooling is
the only variable. We report the out-of-distribution (OOD) area under the ROC curve (AUROC),
recall at a fixed false-alarm rate, and the event-credit mass~\eqref{eq:ecm}.

\section{Results}
\label{sec:results}

\subsection{Safe fast charging under unknown state}
\label{sec:results-main}
\begin{figure}[t]
\centering
\includegraphics[width=\linewidth]{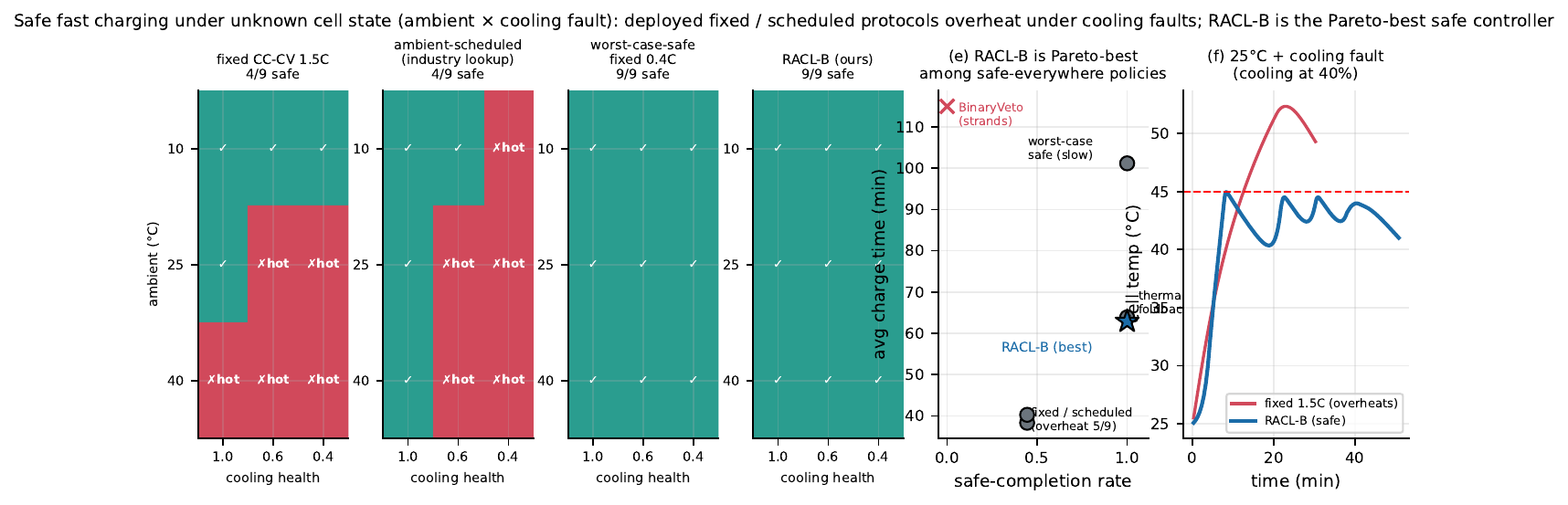}
\caption{Safe fast charging under unknown ambient $\times$ cooling-fault state. Panels show,
for each policy, the $3\times3$ feasibility matrix over ambient temperature (rows) and
cooling health (columns): green/check is safe completion, red/``hot'' is a thermal violation,
grey/dash is an incomplete (stranded) charge. Deployed fixed and ambient-scheduled protocols
are red across the cooling-fault region; RACL-B is feasible everywhere. The right panels show
the safe-completion versus average-time trade-off and a representative trajectory at
$25\degC$ with a cooling fault, where the fixed protocol overshoots while RACL-B rides the
limit.}
\label{fig:envelope}
\end{figure}

Table~\ref{tab:envelope} reports the safe-completion of one deployed setting across the nine
conditions, and Fig.~\ref{fig:envelope} visualizes it. The nominal fixed protocol and the
ambient-scheduled lookup table each overheat in five of the nine conditions: the schedule
adapts to ambient temperature but cannot see the cooling fault, so it overshoots the
$45\degC$ limit (to as high as $52\degC$ at $25\degC$ and $49\degC$ at $40\degC$) whenever
the cooling is degraded. The rigid binary veto never overheats but strands the charge in all
nine conditions, delivering no usable result. The only fixed current that is safe across the
whole envelope must be detuned to $0.4$C and needs $101.2$~min on average. RACL-B safely
completes all nine conditions, reaching $80\%$ SOC in $62.9$~min on average --- $37.9\%$
faster than the worst-case-safe fixed protocol --- with a maximum peak temperature of
$44.96\degC$ under the strict audit.

\begin{table}[t]
\centering
\caption{Safe fast charging under unknown cell state: one deployed setting across the nine
conditions of the envelope~\eqref{eq:envelope} in the DFN + plating model, strict
$45.0\degC$ audit. ``Safe-complete'' counts conditions that reach $80\%$ SOC and stay below
$45\degC$.}
\label{tab:envelope}
\resizebox{\linewidth}{!}{%
\begin{tabular}{lcccc}
\toprule
Policy (one fixed setting) & Safe-complete & Overheats & Strands & Avg.\ time (min) \\
\midrule
CC-CV $1.5$C (nominal)               & 4/9 & 5 & 0 & 38.3 \\
CC-CV ambient-scheduled              & 4/9 & 5 & 0 & 40.2 \\
CC-CV $0.4$C (worst-case-safe)       & 9/9 & 0 & 0 & 101.2 \\
CC-CV + thermal foldback             & 9/9 & 0 & 0 & 63.9 \\
BinaryVeto $3$C                      & 0/9 & 0 & 9 & --- \\
\textbf{RACL-B (ours)}               & \textbf{9/9} & \textbf{0} & \textbf{0} & \textbf{62.9} \\
\bottomrule
\end{tabular}}
\end{table}

Among the policies that are safe across the whole envelope, RACL-B is the most efficient on
both metrics that matter. Table~\ref{tab:strict} reports the strict comparison: RACL-B
reaches $80\%$ SOC fastest and produces the least plated lithium ($14.89$~mAh), because it
repairs the current to the binding margin instead of respecting the worst case throughout.
The advantage over deployed practice is structural rather than a matter of tuning: fixed and
scheduled protocols fail because they commit to a current before they can observe the binding
margin, whereas RACL-B's online repair adapts to an ambient temperature and a cooling health
that it never observes.

\begin{table}[t]
\centering
\caption{Strict comparison among policies that are safe across the whole envelope. RACL-B is
fastest and produces the least plated lithium; all are audited against the same $45.0\degC$
limit.}
\label{tab:strict}
\resizebox{\linewidth}{!}{%
\begin{tabular}{lccc}
\toprule
Safe-everywhere policy & Avg.\ time (min) & Max peak temp.\ ($\degC$) & Avg.\ plated Li (mAh) \\
\midrule
CC-CV $0.4$C (worst-case-safe) & 101.2 & 44.05 & 16.30 \\
CC-CV + thermal foldback       & 63.9  & 44.28 & 15.51 \\
\textbf{RACL-B (ours)}         & \textbf{62.9} & 44.96 & \textbf{14.89} \\
\bottomrule
\end{tabular}}
\end{table}

\subsection{The result is robust to the thermal guard band}
\label{sec:results-robust}
\begin{figure}[t]
\centering
\includegraphics[width=\linewidth]{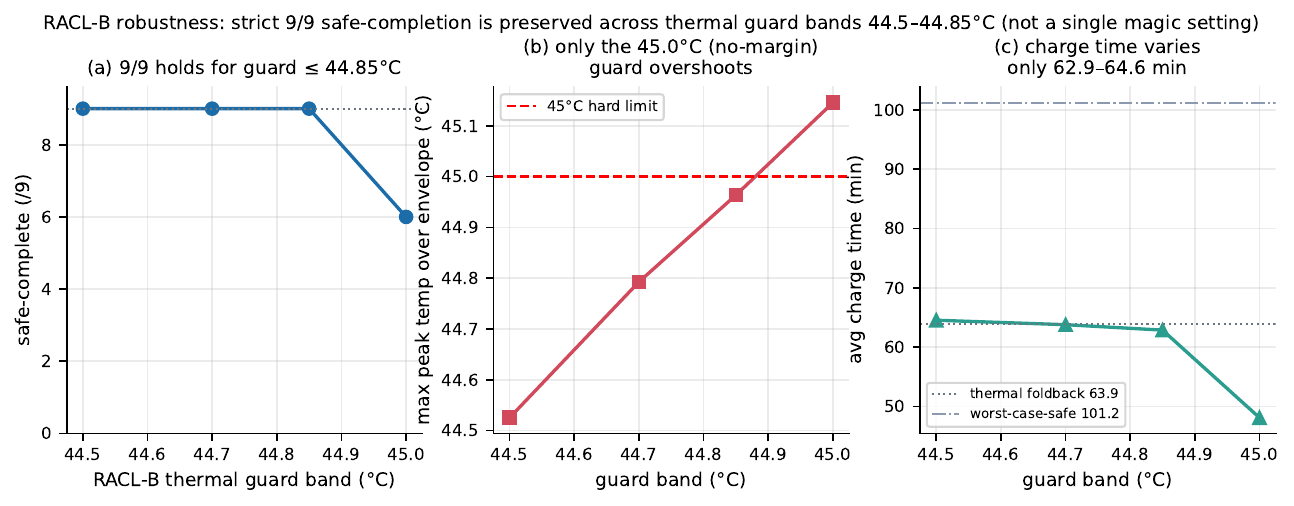}
\caption{Guard-band robustness of RACL-B. Strict $9/9$ safe completion is preserved across
thermal guard bands from $44.5$ to $44.85\degC$ (left); only the zero-margin $45.0\degC$
guard overshoots the hard limit (centre); and the average charge time varies only between
$62.9$ and $64.6$~min, well below the safe fixed and reactive baselines (right).}
\label{fig:guard}
\end{figure}

Because RACL-B's advantage comes from operating close to the thermal limit, we verify that
its safe completion is not a knife-edge tied to one guard-band value. We sweep the control
guard band $T^c$ and audit every run against the same strict $45.0\degC$ limit
(Table~\ref{tab:guard}, Fig.~\ref{fig:guard}). The full $9/9$ safe completion is preserved
for every guard band from $44.5$ to $44.85\degC$, with the average charge time changing only
between $62.9$ and $64.6$~min; only the zero-margin $45.0\degC$ guard allows a discrete-step
overshoot to $45.15\degC$ and loses three conditions. A small thermal margin --- standard
practice in any safety-critical controller --- is therefore sufficient, and the result holds
across a range of settings.

\begin{table}[t]
\centering
\caption{Guard-band sensitivity of RACL-B, audited strictly at $45.0\degC$ over the nine
conditions. The $9/9$ result holds for any guard band from $44.5$ to $44.85\degC$.}
\label{tab:guard}
\resizebox{\linewidth}{!}{%
\begin{tabular}{lccc}
\toprule
RACL-B guard band ($\degC$) & Safe-complete & Max peak temp.\ ($\degC$) & Avg.\ time (min) \\
\midrule
44.5         & 9/9 & 44.53 & 64.6 \\
44.7         & 9/9 & 44.79 & 63.8 \\
44.85 (used) & 9/9 & 44.96 & 62.9 \\
45.0 (no margin) & 6/9 & 45.15 & 48.1 \\
\bottomrule
\end{tabular}}
\end{table}

\subsection{Cross-temperature feasibility}
\label{sec:results-temp}
\begin{figure}[t]
\centering
\includegraphics[width=\linewidth]{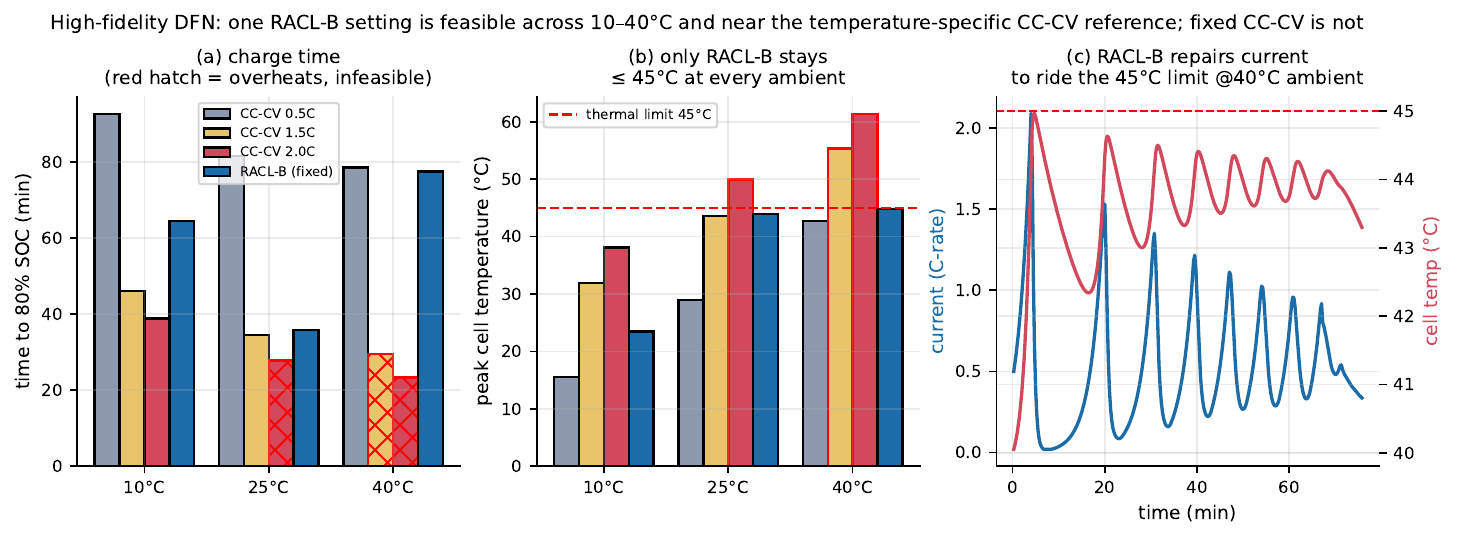}
\caption{Cross-temperature feasibility under the strict audit. Fixed CC-CV protocols either
overheat off-nominal (red, hatched) or must be detuned to a slow rate that is safe
everywhere; RACL-B with one setting stays below $45\degC$ at every ambient (right panel shows
it riding the thermal limit at $40\degC$).}
\label{fig:dfn}
\end{figure}

The same mechanism explains why a single RACL-B setting is feasible across ambient
temperatures where fixed protocols are not (Fig.~\ref{fig:dfn}). Sweeping ambient temperature
at healthy cooling under the strict audit, a fixed $2$C CC-CV overheats above $25\degC$
(reaching $61\degC$ at $40\degC$), and the only fixed rate that is safe across $10$--$40\degC$
is $0.5$C. RACL-B with one setting is feasible at every ambient and is $30\%$, $56\%$, and
$1\%$ faster than that worst-case-safe fixed rate at $10$, $25$, and $40\degC$ respectively,
because it uses the available thermal headroom on cooler cells and tapers only as the cell
approaches the limit.

\subsection{Transient-aware fault monitoring}
\label{sec:results-perception}
\begin{figure}[t]
\centering
\includegraphics[width=\linewidth]{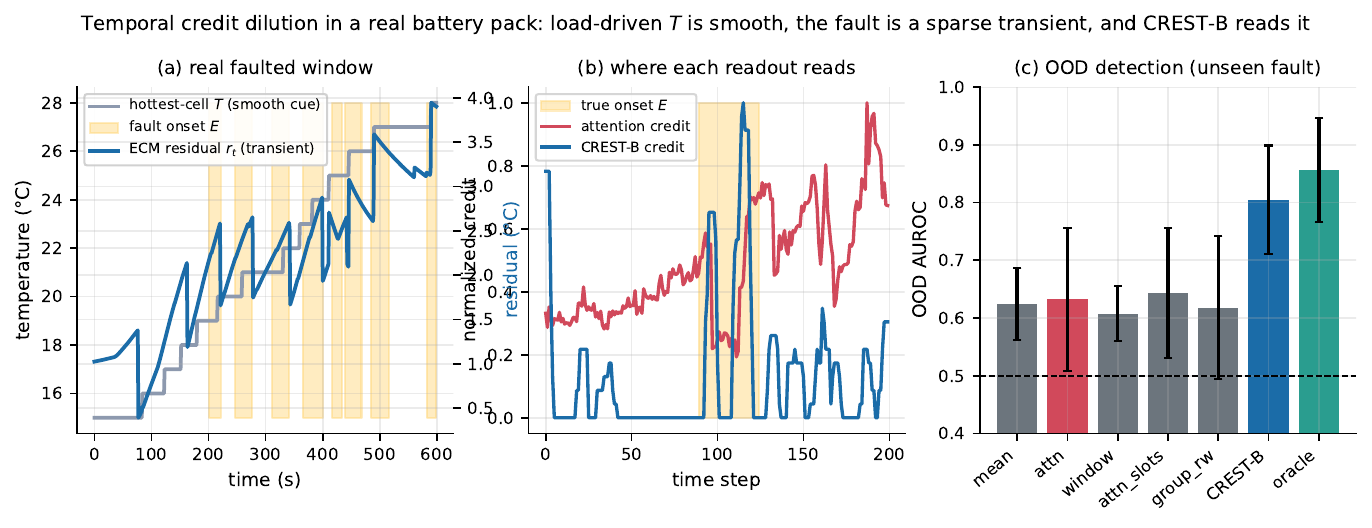}
\caption{Temporal credit dilution in a real faulted window. The load-driven temperature is
smooth while the fault is a brief residual transient (left); attention places its credit on
the smooth trend whereas CREST-B reads the onset (centre); and CREST-B attains the highest
out-of-distribution detection among learned readouts (right).}
\label{fig:teaser}
\end{figure}

\begin{figure}[t]
\centering
\includegraphics[width=\linewidth]{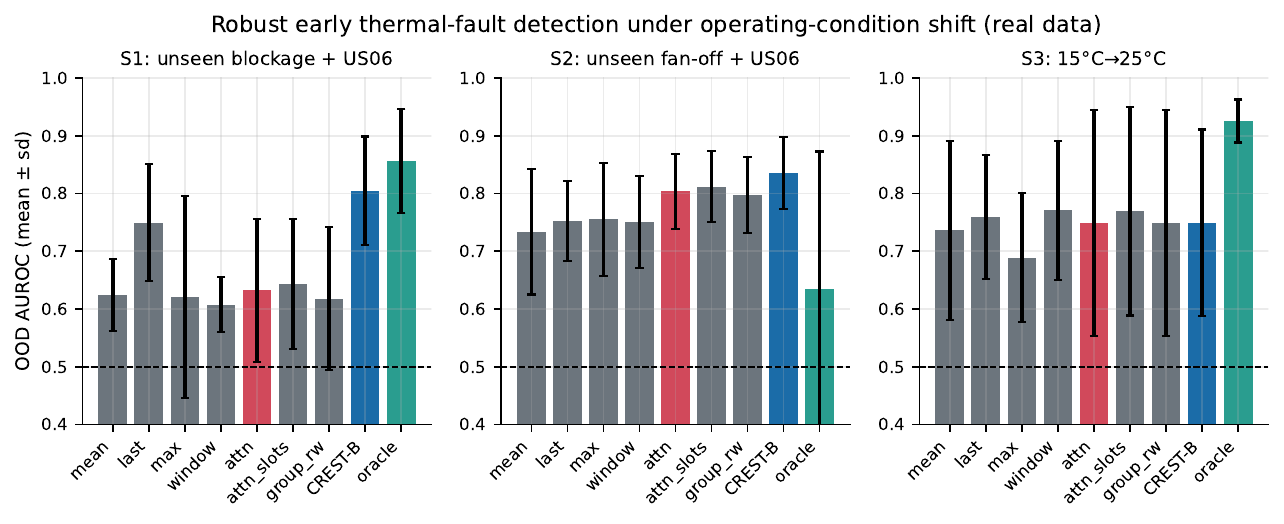}
\caption{Out-of-distribution cooling-fault detection across three operating-condition shifts
on the real introduced-fault pack. CREST-B is the strongest learned readout on the
fault-generalization shifts.}
\label{fig:detect}
\end{figure}

The monitoring side validates the same margin-aware principle on real pack data. On the
introduced-fault dataset, a pooled readout trained with coarse labels attains near-perfect
in-distribution detection yet degrades under operating-condition shift, because it reads the
absolute temperature and load level rather than the residual transient; a hot-but-healthy
aggressive drive cycle is then confused with a genuine cooling fault
(Fig.~\ref{fig:teaser}). Table~\ref{tab:detect} reports the mean OOD-AUROC across the three
shifts. CREST-B is the strongest learned sequence-to-global readout, improving the mean
OOD-AUROC over attention from $0.729$ to $0.796$ and over the best alternative learned
pooling from $0.754$, and approaching the model-based residual reference while remaining a
learned readout that also localizes the onset. On the hardest unseen-fault-type split,
CREST-B improves recall at a $10\%$ false-alarm rate over attention by $4.4\times$, and it
places the highest event-credit mass on the true onset of any learned readout
(Fig.~\ref{fig:detect}), confirming that it reads the physically decisive transient.

\begin{table}[t]
\centering
\caption{Cooling-fault detection under operating-condition shift on the real
introduced-fault pack: mean OOD-AUROC over three shifts. CREST-B is the strongest learned
sequence-to-global readout; the residual threshold is a model-based reference, and the oracle
pools the annotated onset window.}
\label{tab:detect}
\begin{tabular}{lc}
\toprule
Readout (shared per-step encoder) & Mean OOD-AUROC \\
\midrule
Mean pooling          & 0.698 \\
Max pooling           & 0.688 \\
Last-step pooling     & 0.754 \\
Window attention      & 0.709 \\
Attention             & 0.729 \\
Register attention    & 0.741 \\
Group-reweighted attention & 0.722 \\
\textbf{CREST-B (ours)} & \textbf{0.796} \\
\midrule
Residual threshold (model-based reference) & 0.828 \\
Onset-window oracle (upper reference) & 0.805 \\
\bottomrule
\end{tabular}
\end{table}

\subsection{Why pooling fails: theory}
\label{sec:results-theory}
\begin{figure}[t]
\centering
\includegraphics[width=\linewidth]{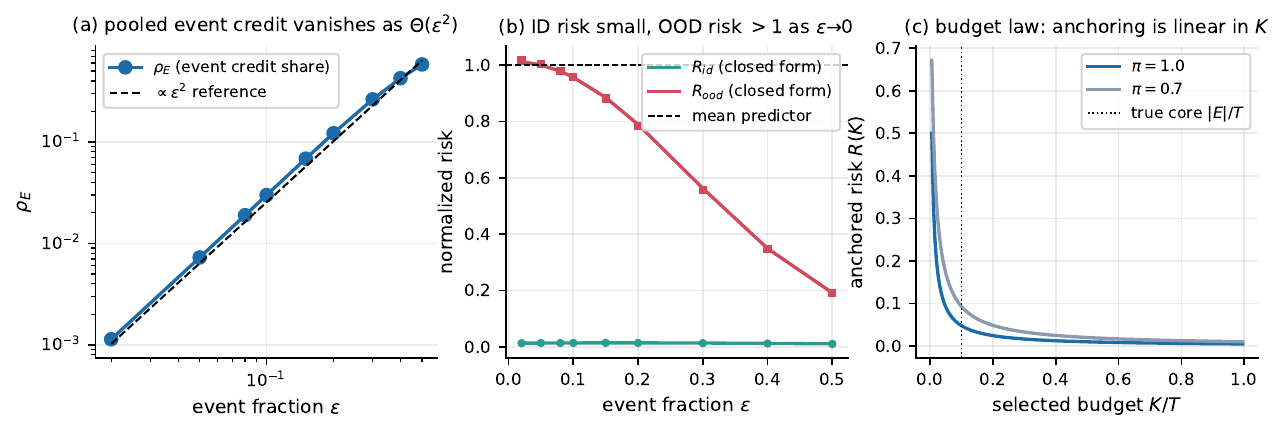}
\caption{Credit-dilution theory. The pooled event credit vanishes as $\Theta(\varepsilon^2)$
(left); the out-of-distribution risk exceeds the mean predictor as the event fraction
$\varepsilon\to0$ while in-distribution risk stays small (centre); and re-anchoring on a
selected core recovers a risk that decreases linearly in the selected budget (right).}
\label{fig:theory}
\end{figure}

The closed-form analysis of Section~\ref{sec:method} is confirmed numerically
(Fig.~\ref{fig:theory}). The pooled event-credit share $\rho_E$ follows the
$\Theta(\varepsilon^2)$ law of~\eqref{eq:rho} --- the measured credit ratio at
$\varepsilon=0.10$ versus $0.02$ is $26.4$, matching the predicted $25.0$ --- and the
out-of-distribution risk of~\eqref{eq:risk} exceeds the mean predictor as $\varepsilon\to0$
(at $\varepsilon=0.02$, $R_{\mathrm{ood}}=1.012>1$) while the in-distribution risk stays
small. Closed-form and Monte-Carlo risks agree to within $0.003$. Pooled readouts therefore
dilute sparse transient evidence by construction, which is exactly what re-anchoring repairs.

\section{Discussion}
\label{sec:discussion}
The results show that the safety of fast charging under uncertainty is not primarily a
question of finding a faster nominal protocol, but of \emph{how the controller responds to a
binding margin it can measure but did not anticipate}. Fixed and ambient-scheduled protocols
encode a commitment made before the cell state is known, and that commitment is unsafe
precisely in the conditions --- a hot cell, a degraded cooling system --- where safety matters
most. Repairing the requested current online to the measured voltage, thermal, and
plating margins turns the same aggressive request into a controller that is safe across the
entire envelope and fast wherever the margins are slack. The $37.9\%$ reduction in charge
time over the only safe fixed protocol, together with the lowest plated lithium and a strict
peak temperature below $45\degC$, is obtained from one deployed setting and is robust across
a range of thermal guard bands.

The monitoring results extend the same principle to perception. A learned BMS monitor that
pools a measurement sequence can place its predictive credit on smooth load and ambient cues
and miss the brief electro-thermal transient that physically signals a cooling fault;
re-anchoring the readout onto that transient makes it the strongest learned monitor for
cross-condition fault localization, and the closed-form analysis explains why the failure is
intrinsic to pooling. Because the perception and control sides treat the same cooling-fault
class through the same margin-aware lens, trusted fault evidence from monitoring can be routed
directly into the repair-before-veto controller, closing the loop from sensing a degraded
margin to acting on it. A high-fidelity electrochemical model with field-resolved transport
and plating provides the safety ground truth for the control results, and a real
introduced-fault pack provides the evidence for the monitoring results.

\section{Conclusion}
\label{sec:conclusion}
We posed safe fast charging as a single-setting, unknown-state problem over a joint envelope
of ambient temperature and cooling-system health, and solved it with a margin-aware
repair-before-veto controller that repairs the requested current online to the tightest
measured voltage, thermal, and lithium-plating margin. In a high-fidelity Doyle--Fuller--Newman
model with lithium plating and a strict $45.0\degC$ audit, the controller safely completes
every condition of the envelope, is $37.9\%$ faster than the only fixed current that is safe
everywhere, produces the least plated lithium, and remains strictly safe across a range of
thermal guard bands, whereas fixed and ambient-scheduled protocols overheat under hidden
cooling degradation and a rigid veto fails to deliver the charge. The same principle yields a
transient-credit monitor that is the strongest learned sequence-to-global readout for
cooling-fault localization on real pack data, supported by a closed-form account of why
pooled readouts dilute transient evidence. Together these give a deployable thermal-safety
guarantee for fast charging and a margin-aware monitor for the same physical fault class.
Future work will integrate the monitor and the controller in a closed hardware loop and
extend the envelope to cell aging and additional fault classes.

\section*{Data and code availability}
The introduced-fault pack dataset is publicly available~\cite{naguib2025data}. The
electrochemical simulations use the open-source PyBaMM framework~\cite{sulzer2021pybamm}.

\bibliographystyle{elsarticle-num}
\bibliography{refs}

\end{document}